# Fluctuating magnetic droplets immersed in a sea of quantum spin liquid

Z. H. Zhu[1,*], B. L. Pan[1,*], L. P. Nie[2,*], J. M. Ni[1,*], Y. X. Yang[1], C. S. Chen[1], Y. Y. Huang[1], E. J. Cheng[1], Y. J. Yu[1], A. D. Hillier[3], X. H. Chen[2,4,5], T. Wu[2,4,5,†], Y. Zhou[6,7,8,†], S. Y. Li[1,4,9,†], & L. Shu[1,4,9,†]

[1]*State Key Laboratory of Surface Physics, and Department of Physics, Fudan University, Shanghai 200433, China*
[2]*CAS Key Laboratory of Strongly-coupled Quantum Matter Physics, Department of Physics, University of Science and Technology of China, Hefei, Anhui 230026, China*
[3]*ISIS Pulsed Neutron and Muon Source, STFC Rutherford Appleton Laboratory, Harwell Campus, Didcot, Oxfordshire OX11 0QX, United Kingdom*
[4]*Collaborative Innovation Center of Advanced Microstructures, Nanjing 210093, China*
[5]*CAS Center for Excellence in Superconducting Electronics (CENSE), Shanghai 200050, China*
[6]*Institute of Physics and Beijing National Laboratory for Condensed Matter Physics, Chinese Academy of Sciences, Beijing 100190, China*
[7]*Songshan Lake Materials Laboratory, Dongguan, Guangdong 523808, China.*
[8]*Kavli Institute for Theoretical Sciences and CAS Center for Excellence in Topological Quantum Computation, University of Chinese Academy of Sciences, Beijing 100190, China.*
[9]*Shanghai Research Center for Quantum Sciences, Shanghai 201315, China*

*These authors contributed equally to this work.
†Corresponding Authors. Email: wutao@ustc.edu.cn (T.W.); yizhou@iphy.ac.cn (Y.Z.); shiyan_li@fudan.edu.cn (S.Y.L.); leishu@fudan.edu.cn (L.S.).

**The search of quantum spin liquid (QSL), an exotic magnetic state with strongly-fluctuating and highly-entangled spins down to zero temperature, is a main theme in current condensed matter physics. However, there is no smoking-gun evidence for deconfined spinons in any QSL candidate so far. The disorders and competing exchange interactions may prevent the formation of an ideal QSL state on frustrated spin lattices. Here we report comprehensive and systematic measurements of the magnetic susceptibility, ultra-low temperature specific heat, muon spin relaxation (µSR), nuclear magnetic resonance (NMR), and thermal conductivity for NaYbSe$_2$ single crystals, in which Yb$^{3+}$ ions with effective spin-1/2 form a perfect triangular lattice. All these complementary techniques find no evidence of long-range**

**magnetic order down to their respective base temperatures. Instead, specific heat, µSR and NMR measurements suggest the coexistence of quasi-static and dynamic spins in NaYbSe$_2$. The scattering from these quasi-static spins may cause the absence of magnetic thermal conductivity. Thus, we propose a scenario of fluctuating ferrimagnetic droplets immersed in a sea of QSL. This may be quite common on the way pursuing an ideal QSL, and provides a brand-new platform to study how a QSL state survives impurities and coexists with other magnetically ordered states.**

Quantum spin liquid (QSL) is a highly-entangled quantum state in which spins remain disordered and dynamic even down to absolute zero temperature due to strong quantum fluctuations[1–6]. Such an exotic state was first proposed from the study of the triangular-lattice Heisenberg antiferromagnets in 1973 by Anderson[1]. Since QSL has potentially tight relationship with high-temperature superconductivity[7] and quantum-information applications[8], it has gained continuous attention in condensed matter physics. The QSL states are characterized by fractional spin excitations, such as spinons, and the detection of these excitations is a crucial issue for identifying QSL in real materials[2–6]. Several QSL candidates have been suggested by experiments, typical examples include triangular-lattice organic compounds $\kappa$-(BEDT-TTF)$_2$Cu$_2$(CN)$_3$[9–11] and EtMe$_3$Sb[Pd(dmit)$_2$]$_2$[12–14], kagome-lattice ZnCu$_3$(OH)$_6$Cl$_2$[15–19], honeycomb lattice $\alpha$-RuCl$_3$[20–23] and H$_3$LiIr$_2$O$_6$ (ref. 24). Despite numerous efforts made in both theoretical and experimental sides, finding realistic "smoking-gun" evidence for QSLs still remains the most challenging task in this field. One of the obstacles comes from the ambiguous role played by the impurities and competing exchange interactions: are they fatal or vital to the survival of a QSL?

In recent few years, the inorganic compound YbMgGaO$_4$, in which Yb$^{3+}$ ions with effective spin-1/2 form a perfect triangular lattice, was argued to have a QSL ground state[25–27]. Although muon spin relaxation (µSR) experiments are consistent with persistent spin dynamics and no static magnetism $\gtrsim 0.003\ \mu_B$ per Yb ion[28,29], the absence of magnetic thermal conductivity at extremely low temperature casts doubts[30], and the observation of frequency-dependent peak of ac magnetic susceptibility suggests a spin-glass ground state in YbMgGaO$_4$[31]. It was argued that the random occupation between Mg$^{2+}$ and Ga$^{3+}$ can mimic a spin-liquid-like state[32]. Thus, a random spin singlet state, or valence bond glass, was proposed to account for the observations[33–35].

Compared with YbMgGaO$_4$, the family of Yb dichalcogenide delafossites NaYb(O, S, Se)$_2$ with effective spin-1/2 is free from the Mg-Ga disorders in non-magnetic layers, thus is of crucial importance to clarify whether there is a QSL ground state in clean triangular lattice of Yb$^{3+}$ ions[36,37]. All three compounds are free from long-range order (LRO) down to 50 mK determined from zero field (ZF) specific heat and µSR measurements[37–40]. Including CsYbSe$_2$ with the same structure, all of them have field-induced magnetic orders[38,41–44]. Very recently, the ground state of NaYbSe$_2$ is claimed to be a QSL with spinon Fermi surface[45]. Pressure-induced superconductivity is also observed in NaYbSe$_2$[46,47], opening up a promising way to study the mechanism of superconductivity in QSL candidates.

Here we report the magnetic susceptibility, specific heat, μSR, nuclear magnetic resonance (NMR), and ultra-low-temperature thermal conductivity measurements on NaYbSe$_2$ single crystals. The absence of magnetic order and spin glass is confirmed by different techniques down to 50 mK. With decreasing temperature in zero field, a hump followed by a linear temperature dependent specific heat is observed. In μSR and NMR measurements, both quasi-static and dynamic spins are found clearly in NaYbSe$_2$. Furthermore, the residual linear term of thermal conductivity at all fields are negligible, pointing to the absence of itinerant fermionic magnetic excitations in NaYbSe$_2$. Our data reveal that NaYbSe$_2$ hosts a ground state of fluctuating ferrimagnetic droplets immersed in a sea of quantum spin liquid on Yb$^{3+}$ triangular lattice.

NaYbSe$_2$ crystallizes in the space group $R\bar{3}m$[36]. As shown in Fig. 1a, in the structure of NaYbSe$_2$, magnetic Yb$^{3+}$ ions form flat triangular layers and each Yb$^{3+}$ ion has 6-fold coordination with O atoms to form an YbO$_6$ octahedron which is edge-sharing with neighboring octahedrons. Interlinked between these flat triangular layers are sheets of pure Na atoms, which structurally removes the widely discussed issue on the site mixing in YbMgGaO$_4$.

The temperature dependences of magnetic susceptibility $\chi$ of NaYbSe$_2$ in external magnetic field $\mu_0H$ = 1 T in two different directions are plotted in Fig. 1b. The absence of magnetic phase transition is confirmed down to 2 K. There is no splitting between zero-field cooling (ZFC) and field cooling (FC) curves of magnetic susceptibility (Fig. S2a), suggesting no spin glass in the system down to 2 K. The inset of Fig. 1b presents a Curie-Weiss (CW) fit with field perpendicular to the $c$ axis. The data above 100 K can be well fitted by CW law, giving effective moment $\mu_{eff}$ = 4.54 $\mu_B$ and the CW temperature $\Theta_{CW}$ = −49.0 K. The value of $\mu_{eff}$ agrees with the theoretical prediction 4.54 $\mu_B$ for trivalent Yb$^{3+}$ ion with $J$ = 7/2. Compared with the CW temperature of YbMgGaO$_4$ (−4 K)[25], $\Theta_{CW}$ is much larger in NaYbSe$_2$. Similar to YbMgGaO$_4$[25], magnetization $M$ remains unsaturated but is smaller up to 7 T at 2 K in NaYbSe$_2$ (Fig. S2b). The larger absolute value of the CW temperature and smaller $M$ indicate stronger AFM interactions in NaYbSe$_2$.

The temperature dependence of specific heat of NaYbSe$_2$ in various fields ($H \parallel c$) from 0.05 to 20 K are shown in Fig. 1c. Consistent with magnetic susceptibility and former reports[43,45], no sharp anomaly of LRO is observed in NaYbSe$_2$. With decreasing temperature, a broad hump of specific heat shows up, whose position shifts to higher temperature in magnetic field. However, we do not observe the field-induced transition peak reported previously, due to the lack of the sufficiently strong field[43]. For the ZF data, after subtracting the contributions from phonon and nuclear Schottky anomaly (Fig. S3), we obtain the magnetic contribution of specific heat $C_{Mag}/T$ as shown in Fig. 2d. As guided by the red dashed line, the temperature independent behavior of $C_{Mag}/T$ below 0.25 K is consistent with a spinon Fermi surface[45]. By integrating $C_{Mag}/T$, we obtain the magnetic entropy $S_{Mag}$, as shown in Fig. 2d. For an effective spin-1/2 system, the theoretical magnetic entropy is $R\ln2$, where $R$ is the gas constant. The residual entropy of NaYbSe$_2$ at 50 mK is only 5.2% of total entropy. Such little entropy remaining suggests low temperature physics is dominated by quantum fluctuations rather than thermal fluctuations, indicating the existence of QSL.

Both μSR and NMR, which measure spin dynamics at different frequency ranges, are powerful tools in clarifying the static and/or dynamic nature of the magnetic ground state. μSR, which uses muon as a probe, is more sensitive to local magnetic field[28,48–51]. As shown in Fig. 2a, the time spectra of muon polarization $P(t)$ in ZF clearly indicates that LRO is absent in ZF-μSR down to 88 mK. The relaxation process of ZF-μSR can be well described by the sum of a Kubo-Toyabe (KT) term and an exponential term:

$$P(t) = f\,G_{KT}(\sigma, t) + (1-f)\exp(-\lambda t), \tag{1}$$

where $f$ is the fraction of the KT term. The fitting function is exactly the same as in the NaYbS$_2$ case[40]. The KT term originates from an isotropic Gaussian distribution of randomly oriented static or quasi-static local fields, whose relaxation rate $\sigma$ is proportional to the root-mean-square (rms) width of the distribution[48]. The exponential term with relaxation rate $\lambda$, originates from dynamic spins. The successful fitting with the above function strongly suggests the coexistence of distinguishable quasi-static spins and dynamic spins.

The temperature dependence of $f$, $\sigma$, and $\lambda$ are plotted in Figs. 2b, and 2c. At high temperature, the value of $f$ is equal to 1, indicating a trivial paramagnetic state. With decreasing temperature below 20 K, $f$ decreases continuously due to the role of magnetic exchange interaction and significant spin dynamic appears which comes to the second term in Eq. (1). Above 10 K, the system remains in a paramagnetic state without quasi-static spins, which is also supported by a temperature-independent NMR intensity (see Fig. S6b and related discussion in Supplementary Materials). Below 6 K, the temperature-dependent $f$ deviates from the descending behavior and shows an upturn behavior, while the temperature-dependent $\sigma$ also shows a clear increasing behavior below 6 K. These results strongly suggest the formation of quasi-static spins at low temperatures. However, both $\sigma$ and $f$ saturates to a finite value at low temperatures, and the saturation value of $f$ indicates that only 23% of the sample becomes quasi-static at base temperature. On the other hand, the temperature-dependent $\lambda$ also exhibits an increasing behavior below 4 K, supporting the enhancement of spin dynamics at low temperatures. The temperature independent behavior of $\lambda$ below 0.2 K suggests the existence of persistent spin dynamics. Additional evidence of the coexistence of quasi-static and dynamic spins in NaYbSe$_2$ comes from longitudinal field (LF) µSR, which yields that the fluctuation rate $v_c$ at 0.1 K is 2.8 MHz (Fig. S4), larger than 1.7 MHz in NaYbS$_2$ (ref. 40).

Similar evidence for the coexistence of quasi-static and dynamic spins is also found in $^{23}$Na NMR experiments. As shown in Fig. 3a, the three-peak structure of $^{23}$Na NMR spectra at high temperature comes from quadrupole splitting of nuclei with spin number $I = 3/2$ (Fig. 3a). With decreasing temperature, a remarkable broadening of line shape occurs below 10 K, suggesting the formation of short-range spin correlations[52]. Meanwhile, the NMR spectra, besides a group of three relatively sharp peaks, start to develop a broad Gaussian background as a new component (Fig. S5a), which should be ascribed to the quasi-static spins as suggested by ZF-µSR. As shown in Fig. 3b, the temperature dependence of full width at half maximum (FWHM) shows a similar increasing behavior for these two components below 10 K, which indicates a close correlation between these two components beyond simple competition. In addition, the quasi-static moment can be estimated from the broad part of NMR spectra, yielding a small value of 0.15 $\mu_B$ (Fig. S6). This result indicates that the quasi-static spins in NaYbSe$_2$ are still fluctuating which is sharp contrast to traditional spin glass[53,54].

Besides NMR spectrum, the nuclear spin-lattice relaxation also supports the coexistence of quasi-static and dynamic spins. An inhomogeneous spin dynamics is indeed observed below 2 K accompanied by the above two-component behavior in spectrum. As shown in Fig. S5c, the stretching exponent $\beta$, which usually depicts the inhomogeneity of spin dynamics, shows a clear decreasing below 2 K with the value well below one. Especially, at the lowest temperature of 0.25 K, there is a clear two-component behavior appearing in the recovery curve of $T_1$ process (Fig. S5b), which is beyond a single $T_1$ fitting with stretching exponent. This is in line with the scenario proposed above with the coexistence of quasi-static and dynamic spins. Finally, the temperature dependence of the spin-lattice relaxation rate $1/T_1$ extracted from the stretched exponential fitting is plotted in Fig. 3c. The broad hump feature around 50 K is usually ascribed to the development of strong spin correlation at low temperature or crystal electric field (CEF) effect[41]. The absence of magnetic order is confirmed again by the absence of any significant critical fluctuation at low temperatures. Below 2 K, $1/T_1$ saturates to a constant, coinciding with the persistent spin dynamics observed in µSR experiments. This result also excludes the possibility of a trivial spin glass phase, suggesting a novel magnetic ground state in NaYbSe$_2$.

To further check the existence of gapless magnetic excitations, we perform the thermal conductivity measurements to probe the possible itinerant excitations. As for a QSL candidate, thermal conductivity measurement is highly advantageous in probing such elementary excitations, since it is only sensitive to itinerant excitations. In a solid, the contributions to thermal conductivity come from various quasi-particles, such as phonons, electrons, magnons, and spinons. Since NaYbSe$_2$ is an insulator, electrons do not contribute to the thermal conductivity at ultra-low temperatures. Additionally, the contribution of magnons can be ruled out due to the absence of magnetic order down to 50 mK. Therefore, thermal conductivity $\kappa$ at ultra-low temperatures can be described by the formula

$$\kappa = aT + bT^\alpha, \qquad (2)$$

where $aT$ and $bT^\alpha$ represent the contribution of possible itinerant gapless fermionic magnetic excitations and phonons, respectively[55,56]. Due to the specular reflections of phonons at the sample surfaces, the power $\alpha$ in the second term is typically between 2 and 3[55,56]. The experiment results in ZF are shown in Fig. 4. The fitting to the data below 0.4 K gives the residual linear term $\kappa_0/T \equiv a = -0.038 \pm 0.007$ mW K$^{-2}$ cm$^{-1}$, and $\alpha = 1.66 \pm 0.04$. This behavior is very similar to YbMgGaO$_4$[30], suggesting the absence of itinerant gapless magnetic excitations (Fig. S7).

We now turn to discuss the ground state of NaYbSe$_2$. The absence of LRO and spin glass in NaYbSe$_2$ is confirmed down to 50 mK. Both μSR and NMR experiments point out that a minority of quasi-static spins and a majority of dynamic spins coexist in NaYbSe$_2$ down to the base temperature. In fact, our specific heat measurements also hint at this picture. The broad hump of $C_{Mag}/T$ around 0.8 K comes from the correlations of quasi-static spins, while the temperature independent behavior below 0.25 K suggests the existence of well-defined magnetic excitations, which is an essential feature of gapless QSLs[24,37,44]. Comparing our results in ZF, these two characteristic temperatures coincides with our μSR data astonishingly. As for the thermal conductivity measurements, the dynamic spins should result in a finite residual linear term $\kappa_0/T$ in NaYbSe$_2$[57–59]. However, the gapless spinons may be strongly scattered by the quasi-static spins, leading to a negligible $\kappa_0/T$.

We propose here a possible picture of mixed state of fluctuating short-range ferrimagnetic droplets and QSL. As for the minority quasi-static spins, there are no long-range but at least short-range correlations between them. They are not static like spin glass, and our NMR result suggests that they are still fluctuating. They only take 23% of the total spins, suggesting they distribute in the system like droplets. When it comes to the dynamic spins, they remain disordered and fluctuating down to our base temperature, exactly matching the definition of QSL. Additionally, there is only less than 5.2% residual entropy at zero temperature, also indicating the presence of

QSL.

Now it brings us to why other methods like magnetic susceptibility and neutron scattering do not observe such ferrimagnetic droplets[36,45]. For magnetic susceptibility technique, it is more sensitive to slower fluctuations, whose limit is about $10^4$ Hz, while NMR and µSR are more sensitive to faster fluctuations. Hence a state could be dynamic in magnetic susceptibility measurements, but quasi-static in NMR and µSR. In inelastic neutron scattering experiments, such fluctuating droplets could also mimic spinon continuum due to its randomness, making it difficult to differentiate.

The ferrimagnetic droplets immersed in a sea of QSL is illustrated in Fig. 5, on which an up-up-down magnetic structure forms within each droplet in accordance with field-induced magnetic orders in $Yb^{3+}$ compounds on triangular lattice[38,41–44]. It is natural to assume that such a ferrimagnetic ordering state has slightly higher energy than the QSL state, which allows the nucleation of ferrimagnetic droplet around a defect. Meanwhile, the thermal fluctuation of these magnetic droplets will give rise to the residual entropy. The ratio between the residual entropy and the total magnetic entropy is estimated to be $\frac{r \ln(1+m/3)}{m \ln 2}$, where $r$ is the volume fraction of droplets and each droplet carries $m$ $Yb^{3+}$ ions. It is expected that the size of the fluctuating droplets will be enhanced by an external magnetic field, resulting in a long ranged up-up-down magnetic order in bulk when the applied magnetic field exceeds some threshold[38,41–44].

In summary, we present specific heat, µSR, NMR, and thermal conductivity measurements on triangular-lattice compound NaYbSe$_2$ single crystals to figure out its ground state. The absence of long-range magnetic order and spin glass is confirmed down to 50 mK. Specific heat, µSR and NMR measurements all find a majority of dynamic spins and a minority of quasi-static spins mixed in NaYbSe$_2$, which is further supported by thermal conductivity measurements. The ground state of NaYbSe$_2$ can be regarded as a mixed state with both QSL and fluctuating short-range ferrimagnetic droplets, providing a platform to study how disorder influence the QSL state.

**References**

1. Anderson, P. W. Resonating valence bonds: A new kind of insulator? *Mater. Res. Bull.* **8**, 153–160 (1973).

2. Balents, L. Spin liquids in frustrated magnets. *Nature* **464**, 199–208 (2010).

3. Savary, L. & Balents, L. Quantum spin liquids: a review. *Reports Prog. Phys.* **80**, 016502 (2017).

**Methods**

**Sample preparation** High-quality NaYbSe$_2$ single crystals were grown by a modified flux method following Schleid and Lissner[59]. Analytically pure Yb powder, Se powder and NaCl as flux in a molar ratio of 2:3:90 were sealed in a quartz tube and heated to 950 °C for 7 days, followed by a maintaining at 950 °C for 7 days. The mixture was slowly cooled down to 600 °C at a rate of 50 °C per day. In the end, reddish black platelets with largest size of 7-8 mm, as shown in the inset of Fig. S1, were separated by dipping in water. The large natural surface was determined to be the (001) plane by X-ray diffraction (XRD), as illustrated in Fig. S1 and no impurity phases were observed, indicating a relatively high crystallization quality.

**Magnetic measurements** The magnetic susceptibility measurements were performed in commercial SQUID and the specific heat was measured in the physical property measurement system (PPMS) (Quantum Design) by the relaxation method.

**μSR measurements** In a μSR experiment, a beam of nearly 100% spin polarized muon is implanted into the sample. Muon spin precesses and relaxes due to inhomogeneous local magnetic field. One can measure the time spectra of muon spin polarization, and the relaxation process can reveal the distribution of local field[27,48,49]. Besides, muon is extremely sensitive to small field, which is a powerful technique to check the essence of magnetic order[50]. ZF and LF μSR measurements were performed down to 88 mK on MuSR spectrometer at ISIS, Rutherford Appleton Laboratory, Chilton, UK. Single crystals of NaYbSe$_2$ were aligned so that the *c* axis was normal to the sample's planar surface and parallel to the initial muon spin polarization, and mounted onto a silver holder covering a circle area of 1 inch in diameter, and 3 mm in thickness. μSR data were analyzed using the MANTID PROJECT and MUSRFIT software package[60]. Subtracting the constant background signal due to silver sample holder, ZF muon spin polarization spectra $P(t)$ can be described by the formula $P(t) = f\, G_{KT}(\sigma, t) + (1 - f)\, \exp(-\lambda t)$, where $G_{KT} = \frac{1}{3} + \frac{2}{3}(1 - \sigma^2 t^2) \exp\left(-\frac{1}{2}\sigma^2 t^2\right)$ is the Kubo-Toyabe (KT) function[47].

**NMR measurements** The $^{23}$Na NMR measurements are taken on one piece of NaYbSe$_2$ single crystal with the mass of 2.6 mg. Because the nuclear gyromagnetic ratios for $^{23}$Na ($\gamma_{Na}$ = 11.2625 MHz/T) is very close to $^{63}$Cu ($\gamma_{Cu}$ = 11.285 MHz/T), we chose the Ag wire to wind NMR coil. In order to define the exact external magnetic field, we fill a small piece of Al foil into the coil. The NMR spectra are obtained by the fast Fourier transformation (FFT) sum of the standard spin-echo signals. The linewidth is extracted from Gauss fitting. Especially, when fitting the spectra of dynamic part between 0.25 K and 1.5 K, we fixed the linewidth of the central line and the satellite line to be the same. The nuclear spin-lattice relaxation rate $1/T_1$ is measured by saturation method below 1.5 K and inverse method for higher temperature. The recovery curve of the nuclear magnetization $M(t)$ is fitted with the function $1 - \frac{M(t)}{M(\infty)} = I_0 \left\{ 0.1 \exp\left[-\left(\frac{t}{T_1}\right)^\beta\right] + 0.9 \exp\left[-\left(\frac{6t}{T_1}\right)^\beta\right] \right\}$. The error bars are determined by the least square method.

**Thermal conductivity measurements** The single crystal selected for the thermal conductivity measurements was a rectangular shape of dimensions $5.38 \times 1.3$ mm$^2$ in the *ab* plane, with a thickness of 0.04 mm along the *c* axis. The thermal conductivity was measured in a dilution refrigerator, using a standard four wire steady-state method with two RuO$_2$ chip thermometers, calibrated *in situ* against a reference RuO$_2$ thermometer. Magnetic fields were applied along the *c* axis for specific heat and thermal conductivity measurements and perpendicular to the heat current in the thermal conductivity measurements.

**Residual entropy** Assuming each ferrimagnetic droplet carries $m$ spin-1/2 ($J_{eff} = 1/2$ local moment) and the fractional volume of the ferrimagnetic droplets is $r$, then the effective spin of each droplet is $\overline{S} = \frac{m}{3} \times \frac{1}{2}$, which gives rise to the upper bound of the residual entropy $R \ln(1 + 2\overline{S}) = R \ln(1 + m/3)$. Thus, the ratio between the residual entropy and the total magnetic entropy at high temperatures has an upper bound of $\frac{r \ln(1+m/3)}{m \ln 2}$.

**Data availability**

All data needed to evaluate the conclusions in the paper are present in the main text or the supplementary information.

**Acknowledgments**

We thank Y. Xu for helpful discussions. This research was funded by the National Natural Science Foundations of China (Grant No. 12034004, No. 11774061, and No. 11774306), the Shanghai Municipal Science and Technology (Major Project Grant No. 2019SHZDZX01 and No. 20ZR1405300), the National Research and Development Program of China, No. 2016YFA0300503, and the Strategic Priority Research Program of Chinese Academy of Sciences (No. XDB28000000).

**Author Contributions**

L.S. and S.Y.L. planned the project. B.L.P. synthesized and characterized the sample. Z.H.Z., Y.X.Y. and C.S.C. carried out the μSR experiments with experimental assistance from A.D.H. L.P.N. carried out the NMR experiments. B.L.P., J.M.N., Y.Y.H., E.J.C. and Y.J.Y. performed the thermal conductivity measurements. L.S., S.Y.L., T.W., Z.H.Z., B.L.P., L.P.N. and J.M.N. analyzed the data. Y.Z. provided the theoretical explanation. L.S., S.Y.L., T.W., Y.Z., X.H.C., Z.H.Z., B.L.P. and L.P.N. wrote the paper.

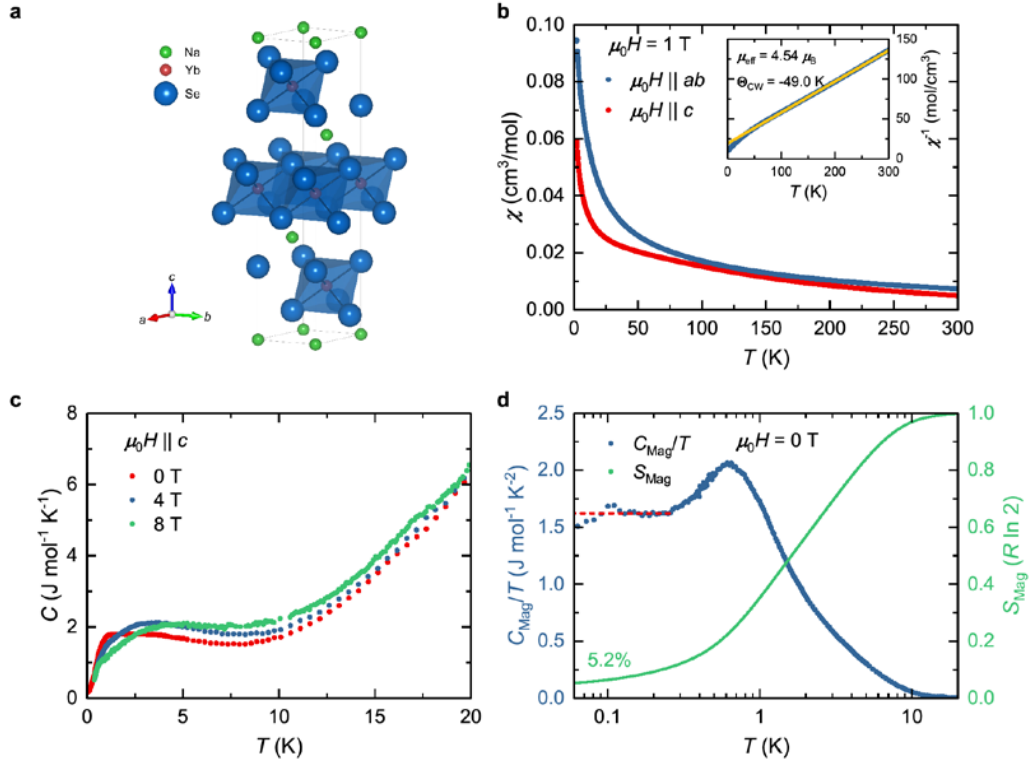

Figure 1: **Basic properties of NaYbSe$_2$.** (a) The unit cell of NaYbSe$_2$. Green spheres: Na. Red spheres: Yb. Blue spheres: Se. (b) The temperature dependence of magnetic susceptibility at $\mu_0 H = 1$ T of NaYbSe$_2$. The inset shows the fitting result of Curie-Weiss law at $T > 100$ K. (c) The temperature dependence of specific heat $C$ at $\mu_0 H = 0, 4, 8$ T. (d) The magnetic specific heat $C_{Mag}/T$ and the calculated magnetic entropy $S_{Mag}$ at zero field. The red dashed line is a guide to eyes to show that $C_{Mag}/T$ is temperature independent at low temperature.

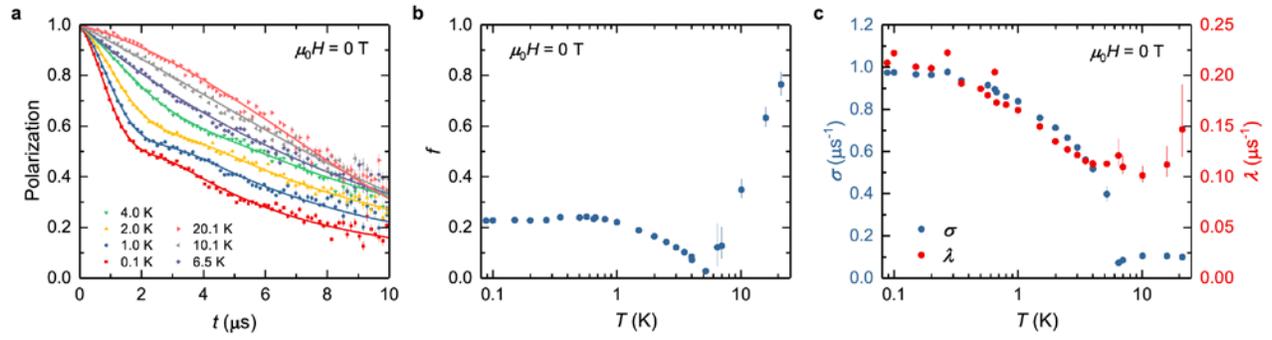

Figure 2: **Zero-field µSR experiment.** (a) Time spectra of ZF-µSR at representative temperatures. The curves are the fittings using Eq. 1. (b) The temperature dependence of fraction of quasi-static spins. (c) The temperature dependence of relaxation rate due to quasi-static and dynamic spins $\sigma$ (blue spheres), and $\lambda$ (red spheres), respectively.

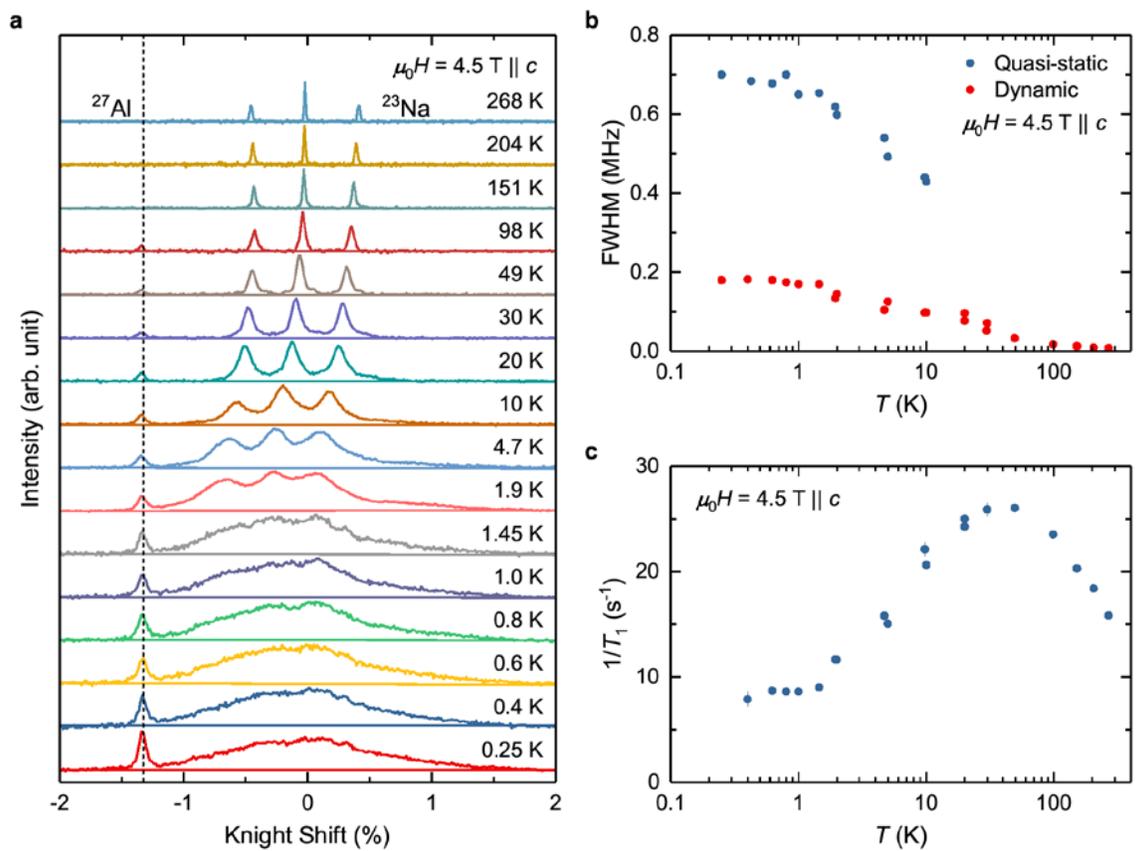

Figure 3: **$^{23}$Na NMR experiments at 4.5 T.** (a) NMR spectra of $^{23}$Na nuclei with external field $\mu_0 H = 4.5$ T parallel to $c$-axis. The sharp peak guided by the black dashed line is the $^{27}$Al peak which is used to calibrate the magnetic field. (b) The temperature dependence of NMR linewidth derived from the spectra as described in SM. The quasi-static (blue spheres) and dynamic (red spheres) components can be easily separated. (c) The temperature dependence of the spin-lattice relaxation rate $1/T_1$.

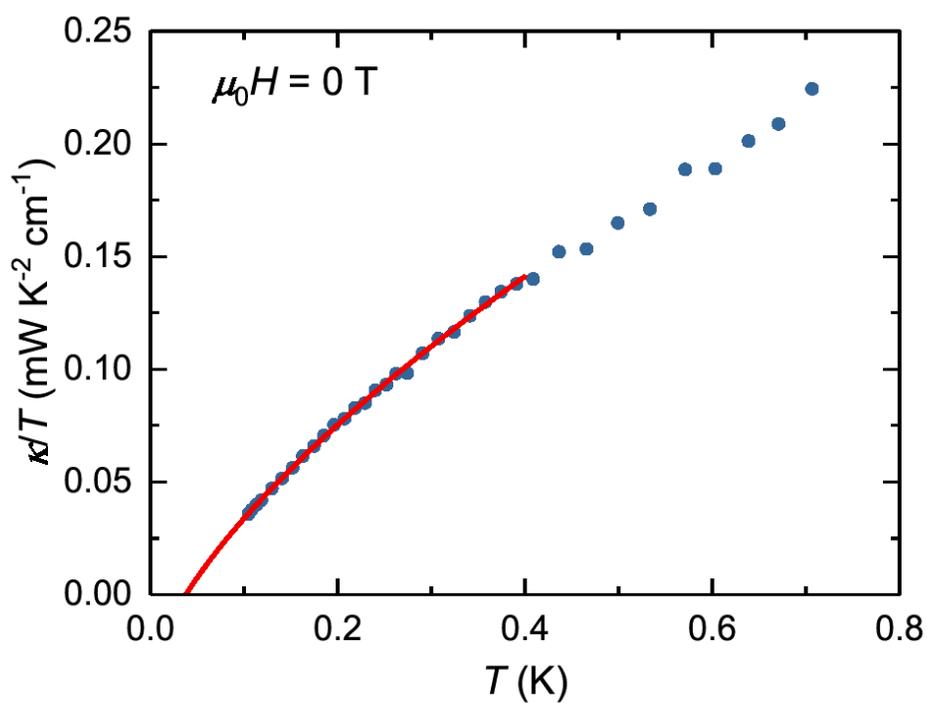

Figure 4: **In-plane thermal conductivity at zero field.** The red solid line is the fit to the data below 0.4 K using Eq. 2.

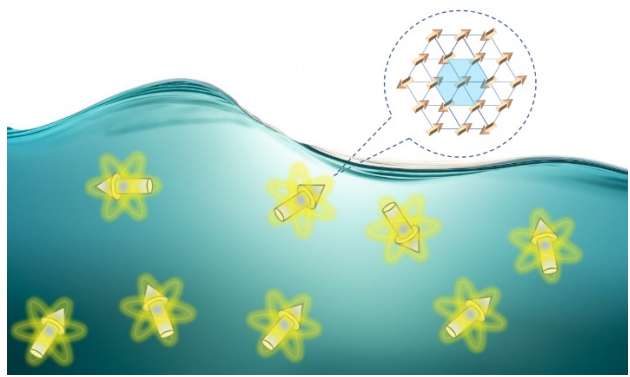

Figure 5: **Magnetic droplets immersed in a sea of quantum spin liquid.** Each droplet has an up-up-down ferrimagnetic structure.

# Supplementary Information for

# Fluctuating magnetic droplets immersed in a sea of quantum spin liquid

Z. H. Zhu[1,*], B. L. Pan[1,*], L. P. Nie[2,*], J. M. Ni[1,*], Y. X. Yang[1], C. S. Chen[1], Y. Y. Huang[1], E. J. Cheng[1], Y. J. Yu[1], A. D. Hillier[3], X. H. Chen[2,4,5], T. Wu[2,4,5,†], Y. Zhou[6,7,8,†], S. Y. Li[1,4,9,†], & L. Shu[1,4,9,†]

[1]*State Key Laboratory of Surface Physics, and Department of Physics, Fudan University, Shanghai 200433, China*

[2]*CAS Key Laboratory of Strongly-coupled Quantum Matter Physics, Department of Physics, University of Science and Technology of China, Hefei, Anhui 230026, China*

[3]*ISIS Pulsed Neutron and Muon Source, STFC Rutherford Appleton Laboratory, Harwell Campus, Didcot, Oxfordshire OX11 0QX, United Kingdom*

[4]*Collaborative Innovation Center of Advanced Microstructures, Nanjing 210093, China*

[5]*CAS Center for Excellence in Superconducting Electronics (CENSE), Shanghai 200050, China*

[6]*Institute of Physics and Beijing National Laboratory for Condensed Matter Physics, Chinese Academy of Sciences, Beijing 100190, China*

[7]*Songshan Lake Materials Laboratory, Dongguan, Guangdong 523808, China.*

[8]*Kavli Institute for Theoretical Sciences and CAS Center for Excellence in Topological Quantum Computation, University of Chinese Academy of Sciences, Beijing 100190, China.*

[9]*Shanghai Research Center for Quantum Sciences, Shanghai 201315, China*

*These authors contributed equally to this work.

†Corresponding Authors. Email: wutao@ustc.edu.cn (T.W.); yizhou@iphy.ac.cn (Y.Z.); shiyan_li@fudan.edu.cn (S.Y.L.); leishu@fudan.edu.cn (L.S.).

## 1  X-ray diffraction

Room-temperature X-ray diffraction pattern from the largest natural surface of single crystalline NaYbSe$_2$ is plotted in Fig. S1. Only (00$l$) peaks are observed, demonstrating that the largest natural surface is (001) plane. A photo of NaYbSe$_2$ single crystal is also shown in the inset of Fig. S1.

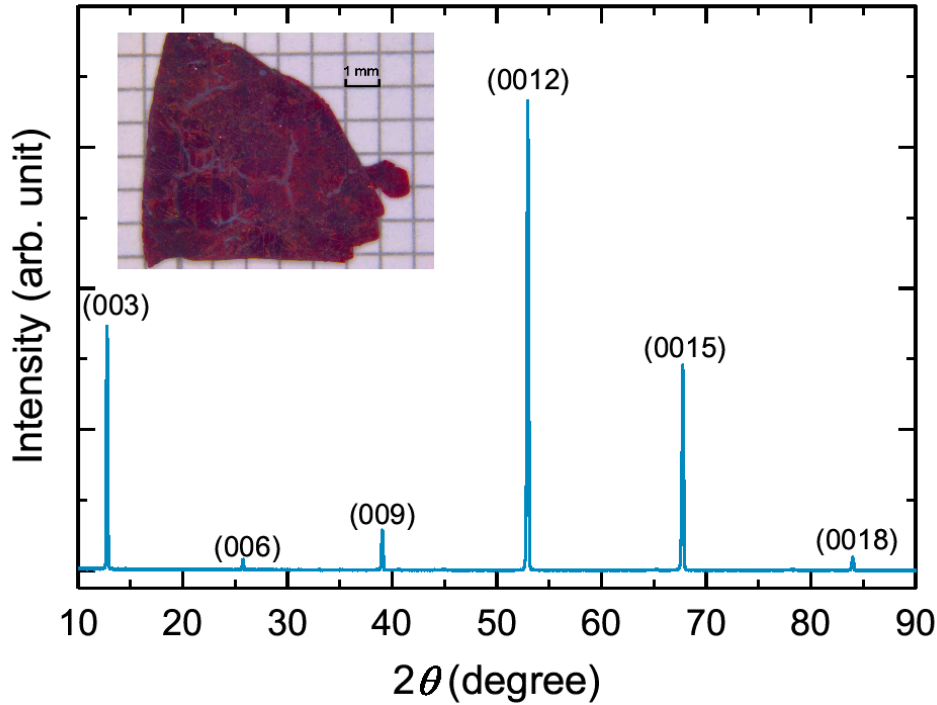

Figure S1: X-ray diffraction pattern of single crystalline NaYbSe$_2$. Inset: photograph of a NaYbSe$_2$ single crystal.

## 2  Magnetic properties

To check if there is any evidence of spin glass, we performed zero-field cooling (ZFC) and field cooling (FC) measurements of magnetic susceptibility. As plotted in Fig. S2a, in 1 T external field, no splitting between ZFC and FC curves of magnetic susceptibility is observed with field parallel and perpendicular to the *ab*-plane, indicating the absence of spin glass.

We also measured the field dependence of magnetization at 2 K with field parallel and perpendicular to the *ab*-plane. Both curves show no sign of saturation up to 7 T. Although it was reported that a 4 T field in the *ab*-plane can induce magnetic order[1], the transition temperature is about 1 K which is below the temperature range of our experiment.

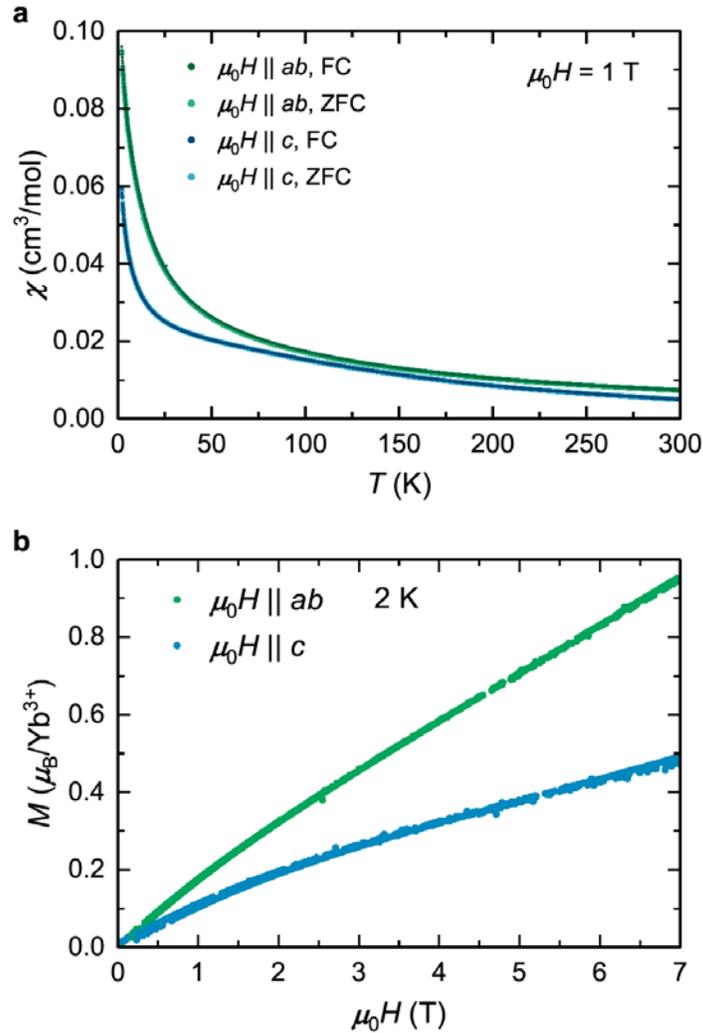

Figure S2: Magnetic properties of NaYbSe$_2$. (a) The temperature dependence of magnetic susceptibility at $\mu_0 H$ = 1 T ($\parallel ab$ and $\parallel c$) with FC and ZFC. (b) The field dependence of the magnetization at 2 K with $\mu_0 H \parallel ab$ and $\parallel c$.

## 3   Specific heat

The analysis process of the specific heat $C$ in zero field is shown in Fig. S3. First, we subtract the lattice contribution from the total specific heat as shown in Fig. S3a. The lattice contribution is obtained from the specific heat result from the work of Dai *et al.*[2].

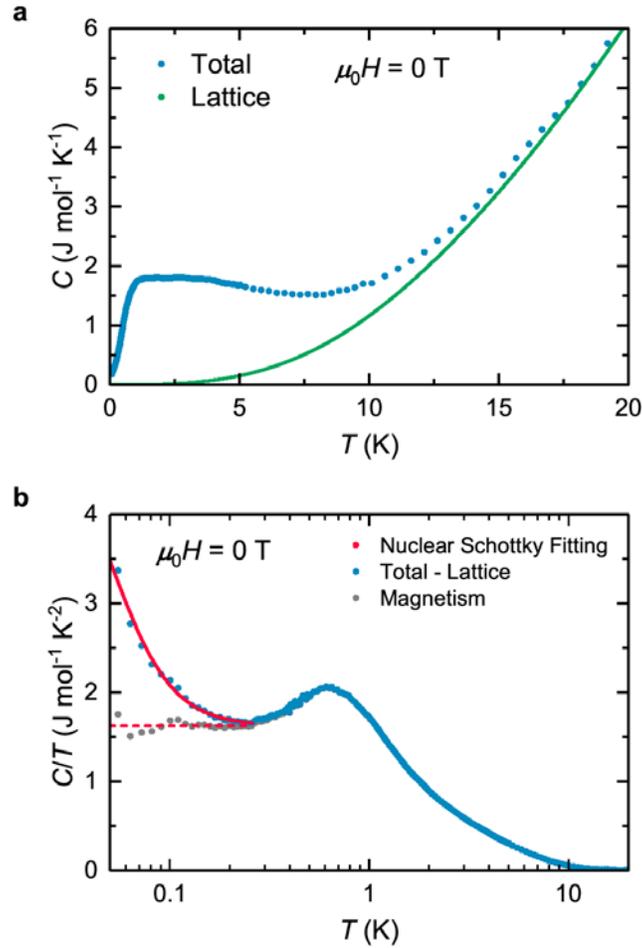

Figure S3: Specific heat results of NaYbSe$_2$. (a) The temperature dependence of the total and lattice specific heat in zero field. (b) The fitting of the nuclear Schottky anomaly. Blue spheres: total specific heat subtracting the lattice contribution. Red curve: the fitting curve using Eq. S1. Red dashed line: the $T$-linear term in the fitting function. Grey spheres: the magnetic contribution as shown in Fig. 1d in the main text.

We now consider the ultralow-temperature specific heat. Based on the work of Dai *et al.*[2], a *T*-linear specific heat is expected after subtracting nuclear Schottky anomaly contribution. We fit the data below 0.25 K using a *T*-linear term plus a two-state Schottky anomaly term:

$$C = \gamma T + nC_{\text{Schottky}}, \tag{S1}$$

where

$$C_{\text{Schottky}} = R\frac{g_0}{g_1}\left(\frac{\Delta}{k_B T}\right)^2 \frac{\exp(\Delta/k_B T)}{\left(1+\frac{g_0}{g_1}\exp\left(\frac{\Delta}{k_B T}\right)\right)^2}. \tag{S2}$$

Here, $R$ is the gas constant, and $k_B$ is the Boltzmann constant. The fitting result is shown as the red solid curve in Fig. S3b. The fitting yields $\gamma = 1.62$ J mol$^{-1}$ K$^{-2}$ shown as the red horizontal dashed line, the value of which is close to previous report[2]. The area between the red solid curve and dashed line represents the Schottky anomaly term. In the Schottky anomaly term, $g_0/g_1$ is the ratio between degeneracies of the ground and excited state, which is set to 1 to simplify the fitting, $n = 0.026$ is the fraction of atoms that induce nuclear Schottky anomaly, and $\Delta = 9.03$ $\mu$eV is the energy difference between the two states, respectively. Subtracting the obtained nuclear Schottky anomaly term, we get the magnetic contribution $C_{\text{Mag}}/T$ as shown in Fig. 1d in the main text or the grey spheres shown in Fig. S3b. The calculated magnetic entropy $S_{\text{Mag}}$ shown in Fig. 1d is derived by integrating $C_{\text{Mag}}/T$. We assume that the spins are totally free above 20 K, meaning that $S_{\text{Mag}}$ is $R\ln2$ at 20 K, and then derive the residual entropy at base temperature is 5.2% of $R\ln2$.

## 4 Longitudinal field muon spin relaxation

Representative longitudinal field $\mu$SR (LF-$\mu$SR) at 0.1 K in various fields $\mu_0H$ along *c*-axis are shown in Fig. S4a. In LF-$\mu$SR experiments, with field increasing, the muon spin is gradually decoupled from the static and quasi-static local field, and affected by the external field[3]. In the time spectra, this effect displays as a suppression of the relaxation. But the relaxation due to dynamic field is much more robust. Thus we can differentiate the quasi-static and dynamic field. The constant background derived from the fitting of ZF-$\mu$SR is also subtracted. The LF polarization spectra $P(t)$ can be well described by the formula

$$P(t) = fG_{\text{LF}}^{\text{KT}}(\mu_{0H}, \sigma, t) + (1-f)e^{-\lambda t}, \tag{S3}$$

where $f$ and $\sigma$ are fixed at the value derived from the fitting of ZF-$\mu$SR at 0.1 K, and

$$G_{LF}^{KT}(\mu_0 H, \sigma, t) = 1 - \frac{2\sigma^2}{(\gamma_\mu \mu_0 H)^2}\left[1 - e^{-\frac{1}{2}\sigma^2 t^2} \cos(\gamma_\mu \mu_0 H t)\right]$$

$$+ \frac{2\sigma^4}{(\gamma_\mu \mu_0 H)^3} \int_0^t e^{-\frac{1}{2}\sigma^2 \tau^2} \sin(\gamma_\mu \mu_0 H \tau) \, d\tau, \tag{S4}$$

is the static LF-KT function, where $\gamma_\mu/2\pi = 135.54$ MHz/T is the gyromagnetic ratio of muon[3]. Since all other parameters are fixed, the only derived parameter is the dynamic relaxation rate $\lambda$. The field dependence of $\lambda$ at 0.1 K is shown in Fig. S4b. Normally, the field dependence of $\lambda$ can be described by the conventional Redfield formula which supposes the time correlation function of dynamic spin takes a simple exponential Markovian form[4]

$$\lambda = \frac{(\gamma_\mu B_{loc}^{rms})^2 \tau_c}{1 + (\gamma_\mu \mu_0 H \tau_c)^2}, \tag{S5}$$

where $B_{loc}^{rms}$ is the root-mean-square (rms) of local fields, and $\tau_c$ is the correlation time. However, similar to YbMgGaO$_4$ and NaYbS$_2$, the field dependence of $\lambda$ cannot be described by the Redifield formula[5,6]. Instead, the correlation function takes a more general form $S(t) \sim (\tau/t)^x \exp(-\nu t)$, where $\tau$ and $1/\nu$ are the early and late time cutoffs, respectively[4]. When $x = 0$, the correlation function degenerates back into the Markovian case. Such a more general correlation function yields[4]

$$\lambda(H) = 2\Delta^2 \tau^x \int_0^\infty t^{-x} \exp(-\nu t) \cos(\gamma_\mu \mu_0 H t) \, dt \tag{S6}$$

As shown in Fig. S4b, the fitting of the field dependence of $\lambda$ to Eq. S6 gives $x = 0.39$ and $\nu = 2.8$ MHz, which is close to the NaYbS$_2$ case ($x = 0.44$ and $\nu = 1.7$ MHz)[6], and comparable to YbMgGaO$_4$ ($x = 0.66$ and $\nu = 9.4$ MHz)[5].

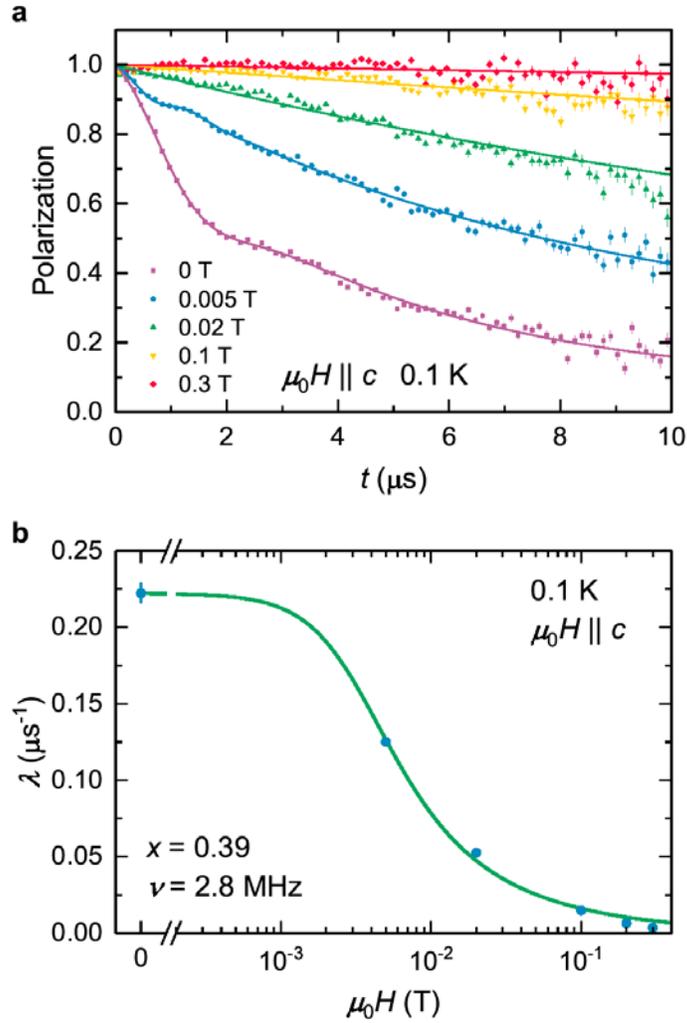

Figure S4: LF-$\mu$SR results at 0.1 K with $\mu_0H$ parallel to $c$-axis. (a) Time spectra of polarization of LF-$\mu$SR at 0.1 K. The curves are the fits to the data with Eq. S4. (b) Field dependence of the dynamic relaxation rate $\lambda$ at 0.1 K. The red curve is the fit with Eq. S6.

## 5  Nuclear magnetic resonance

The fitting details to the $^{23}$Na NMR spectra below 10 K are shown in Fig. S5a. The spectra at low temperatures can be decomposed into two components. One is a broad Gaussian background owing to the quasi-static spins, and the other is three narrow peaks which stand for the dynamic spins. The fittings to the recovery curve are plotted in Fig. S5b. The most of recovery curves at different temperatures can be fitted by a standard $T_1$ formula with a stretching exponent $\beta$ which usually depicts the inhomogeneity of spin dynamics[8]. The temperature dependence of the stretching exponent $\beta$ is shown in Fig. S5c. The decrease of $\beta$ below 2 K suggests a strong inhomogeneity in NaYbSe$_2$. It should be noted that, with the continuous decrease of $\beta$, two relaxation processes with different relaxation rates can be observed at the lowest temperature of 0.25 K, supporting a coexistence of quasi-static and dynamic components. The blue and green dashed curves are the quasi-static and dynamic components, respectively. In addition, the broad hump around 50 K in $1/T_1$ could be the indicator for the development of strong spin correlation at low temperatures or the crystal electric field (CEF) effect in NaYbSe$_2$[7].

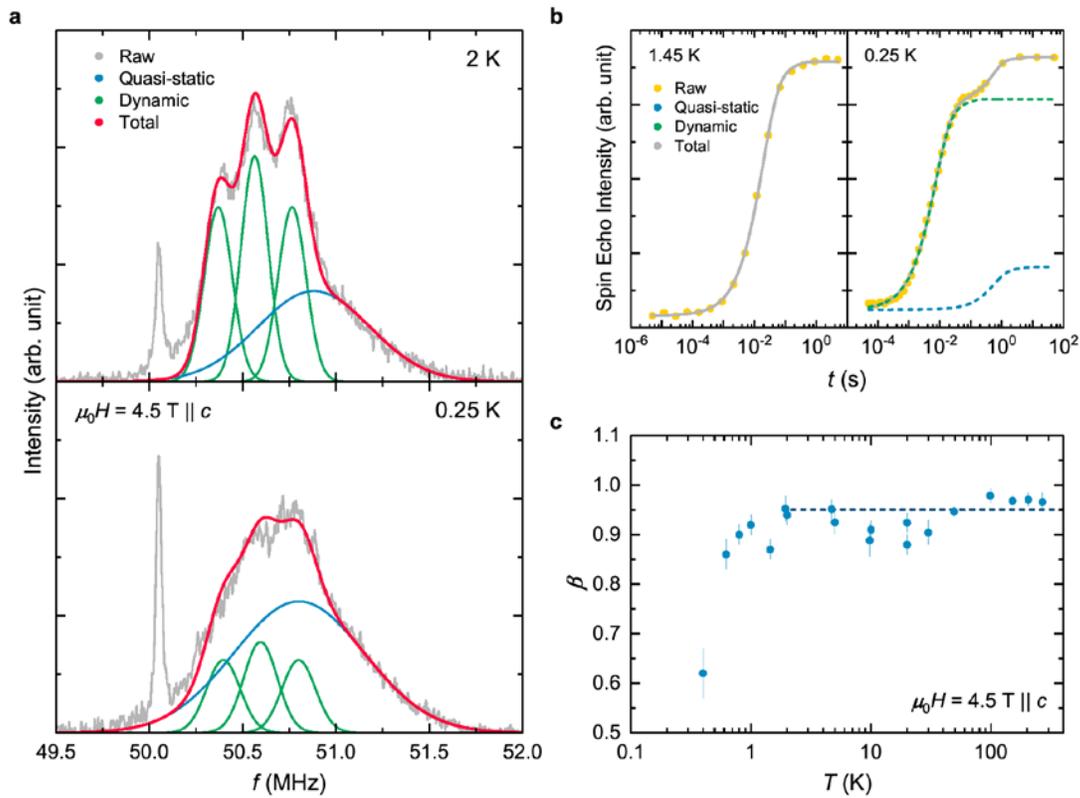

Figure S5: The details for the fitting of NMR spectra and recovery curves. (a) Fitting of the $^{23}$Na spectra with two components at low temperatures. Three green narrow peaks stand for the dynamic component, and the blue broad peak represents the quasi-static component. The red curve is the total fitting to the raw data. (b) The recovery curve of nuclear magnetization. At 0.25 K, a remarkable two-component behavior appears, which is also ascribed to the coexistence of quasi-

static and dynamic spins. The blue and green dashed curves represent the quasi-static and dynamic components in the total fitting. (c) The temperature dependence of the stretching exponent $\beta$.

To further figure out the nature of the observed quasi-static spins, we have analyzed the Knight shift in more details. In single spin component system, the Knight shift is proportional to the bulk magnetic susceptibility $\chi$:

$$K(T) \sim A_s \chi(T) \tag{S7}$$

In this case, $K - \chi$ plot can be used to estimate the magnetic hyperfine coupling tensor $A_s$. As shown in Fig. S6a, by analyzing the $K - \chi$ plot, we find that the value of $A_s$ is estimated to be $-0.095$ T/$\mu_B$ in the temperature range from 268 and 98 K. Then the value of $A_s$ changes to $-0.4115$ T/$\mu_B$ in the temperature range from 98 and 5 K. The change of $A_s$ around 98 K can be ascribed to the CEF effect[7]. After knowing the value of $A_s$, we try to estimate the magnitude of the quasi-static spins. The line width of the quasi-static component at 0.25 K is about 0.7 MHz. Using the gyromagnetic ratio $\gamma_{Na}$ = 11.2625 MHz/T, we can get the distribution of internal field at Na site is about 0.062 T. Such a distribution of internal field should come from the quasi-static component of Yb$^{3+}$ ion . Hence we can estimate that the moment of the quasi-static component is about 0.15 $\mu_B$, which is much smaller than $\mu_{eff}$ = 0.5 $\mu_B$ with $H \parallel c$ estimated from low-temperature magnetization curves. This result indicates the fluctuating nature of the quasi-static spins.

It should be noted that the wipe-out effect in NMR spectrum might lead to an underestimation of the line width. To verify the wipe-out effect on line width, the temperature-dependent integral area of $^{23}$Na NMR full spectra multiplied by temperature is plotted in Fig. S6b. Usually, if there is no wipe-out effect or intensity loss, it means that the magnetic system should be still in a paramagnetic state. As shown in Fig.S6b, there is no detectable wipe-out effect above 2 K, confirming our conclusion on a trivial paramagnetic state above 6 K from our $\mu$SR results (Fig. 2). Below 2 K, a clear wipe-out effect appears before taking consideration of the $T_2$ effect. In order to further check the $T_2$ effect, we have measured $T_2$ at the frequency corresponding to the central line of the dynamic component at 0.25 K. As shown by the red point in Fig. S6b, although the precise measurement on $T_2$ is not trivial for an inhomogeneous system, it is quite clear that the intensity loss is mainly caused by the $T_2$ effect instead of static spin order. Anyway, the limited wipe-out effect rules out a conventional spin glass in NaYbSe$_2$.

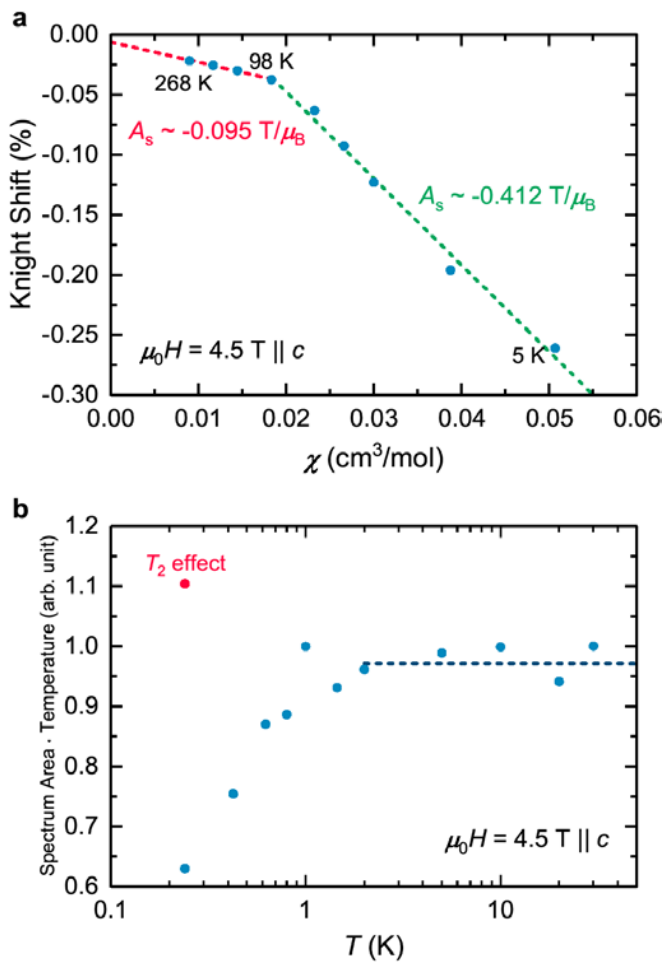

Figure S6: Analysis on hyperfine coupling tensor and wipe out effect. (a) The $K - \chi$ plot. The red dash line is the fitting for hyperfine coupling tensor by using the data between 268 and 98 K. The green dash line is the fitting for hyperfine coupling tensor by using the data between 98 and 5 K. (b) The temperature dependence of the integral area of $^{23}$Na NMR full spectra multiplied by temperature. The value is normalized to that at 30 K. The red point the revised data point at 0.25 K by considering the correction of $T_2$ effect.

## 6  Thermal conductivity

The temperature dependence of in-plane thermal conductivity $\kappa/T$ in various fields ($\mu_0 H \parallel c$) are shown in Fig. S7a. The residual linear term $\kappa_0/T$ is virtually zero in all fields, indicating the absence of itinerant gapless magnetic excitations. Fig. S7b plots the field dependence of the $\kappa/T$ at 0.2, 0.3, and 0.4 K. For $\mu_0 H < 3$ T, $\kappa/T$ is independent of fields. With increasing fields, the spins are increasingly polarized, thus reducing the scattering of phonon, leading to the rapid enhancement of thermal conductivity from 3 to 5 T.

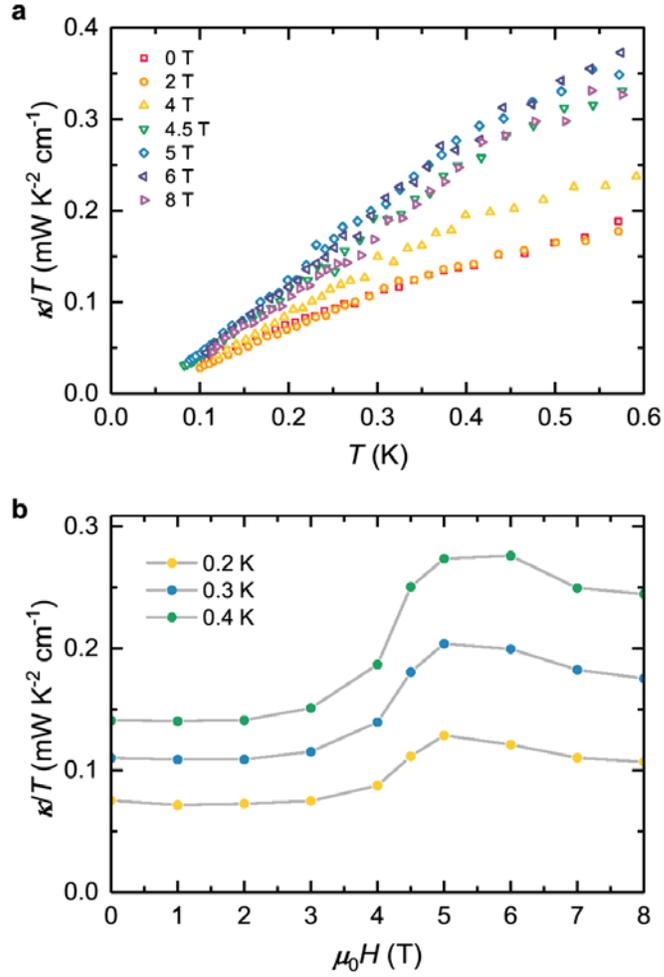

Figure S7: In-plane thermal conductivity of NaYbSe$_2$. (a) In-plane thermal conductivity of NaYbSe$_2$ at various fields ($\mu_0 H \parallel c$). (b) The field dependence of $\kappa/T$ at 0.2, 0.3, and 0.4 K.